\documentclass[twocolumn,epsfig]{revtex4}

\usepackage{graphicx}
\usepackage{amssymb}
\usepackage{color}

\newcommand{\figsize}{8.3cm}

\begin{document}

\draft
\title{Abnormal high-$Q$ modes of coupled stadium-shaped microcavities}
\author{Jung-Wan Ryu,$^{1}$}
\email{jungwanryu@gmail.com}
\author{Soo-Young Lee,$^2$ Inbo Kim,$^2$ Muhan Choi,$^2$ Martina Hentschel,$^{3}$ and Sang Wook Kim$^{4}$}
\affiliation{$^1$Department of Physics, Pusan National University, Busan 609-735, South Korea\\
$^2$School of Electronics Engineering, Kyungpook National University, Daegu 702-701, Korea\\
$^3$Institut f\"ur Physik, Technische Universit\"at Ilmenau, D-98684 Ilmenau, Germany\\
$^4$Department of Physics Education, Pusan National University, Busan 609-735, South Korea
}

\begin{abstract}
It is well known that the strongly deformed microcavity with fully chaotic ray dynamics cannot support high-$Q$ modes
due to its fast chaotic diffusion to the critical line of refractive emission.
Here, we investigate how the $Q$ factor is modified when two chaotic cavities are coupled, and show that some modes,
whose $Q$ factor is about 10 times higher than that of the corresponding single cavity, can exist.
These abnormal high-$Q$ modes are the result of an optimal combination of coupling and cavity geometry.
As an example, in the coupled stadium-shaped microcavities, the mode pattern extends over both cavities
such that it follows a whispering-gallery-type mode at both ends, whereas a big coupling spot forms at the closest contact of the two microcavities.
The pattern of such a 'rounded bow tie' mode allows the mode to have a high-$Q$ factor. This mode pattern minimizes the leakage of light at both ends
of the microcavities as the pattern at both ends is similar to whispering gallery mode.
\end{abstract}
\maketitle
\narrowtext

Microdisk lasers have ultra-low lasing threshold since the whispering gallery modes (WGMs) excited have high-$Q$ factors
due to the strong light confinement by total internal reflection \cite{McC92,Yam93,Vah03}.
Recently, coupled two and multiple microdisk resonators have been studied in the context of photonic molecules \cite{Bay98,Bor06}
and coupled resonator optical waveguides (CROWs) \cite{Yar99}.
On the other hand, the asymmetric microcavities deformed from microdisk have attracted much attention because they exhibit directional emission
as the rotational symmetry is broken and also provide an analog of open quantum system with chaotic dynamics \cite{Noe97,Gma98}.
Although many works have been devoted to single deformed microcavities, the coupled deformed microcavity resonators have not been thoroughly investigated.

In micro-lasers the high-$Q$ modes are  activated at low input current, meaning a low lasing threshold \cite{Sie86}.
The lasing threshold of a conventional laser of Fabry-Perot type resonator is inversely proportional to the size of the laser
and then it is not easy to make small-size laser with low lasing threshold.
It is known in microcavity lasers that the lasing threshold
is determined by how high $Q$ modes are distributed near gain center of the laser \cite{Har03,Har05}.
Various mode characteristics of chaotic microcavities, such as $Q$-factors and mode patterns,
have been investigated and it is pointed out that most modes of a low-refractive-index chaotic microcavity exhibit localization
along the simple periodic orbits \cite{Leb06,Fan07,Lee09,Ryu13}.

In this Letter, we study mode properties, especially $Q$ factor, of coupled chaotic microcavities,
and find unexpected high-$Q$ mode in coupled stadium-shaped microcavities,
its $Q$ value is about 10 times higher than that of single stadium-shaped one.
Such abnormal high-$Q$ mode shows localization along a 'rounded bow tie' pattern which exhibits WGM-like patterns
in the circular parts of both ends and rather wide interference pattern near the interspace between two microcavities, cf. Fig.~\ref{fig2} (b).
The WGM-like pattern at both ends implies small evanescent leakage.
We emphasize that the WGM-like pattern cannot be formed in single stadium-shaped microcavity with chaotic dynamics.
It is the coupling of two {\em chaotic} microcavities that allows the high $Q$ mode to take place.

The wavenumber $k$ of the resonances and the corresponding mode patterns of a microcavity can be obtained by solving the Helmholtz equation,
$\left[\nabla^2+n^2(\mathbf{r})k^2\right]\psi = 0$, where $n$ is the refractive index, by using the boundary element method \cite{Wie03}.
Here, as a chaotic microcavity, we take the Bunimovich stadium \cite{Bun79}-shaped microcavity with $n=1.45$ (see insets in Fig.~1),
which consists of two half circles of radius $R$ and linear segments of length $L=2R$.
Once the complex wavenumber $k$ is numerically obtained, the $Q$-factor of the corresponding mode
is given by $Q=-\mathrm{Re}(kR)/2\mathrm{Im}(kR)$.

\begin{figure}[tb]
\centerline{\includegraphics[width=\figsize]{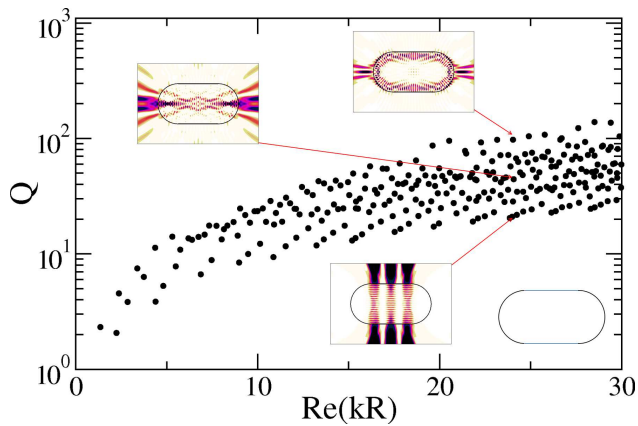}}
\caption{(color online) $Q$ factors of resonant modes in a single stadium-shaped microcavity with $n=1.45$.
Three selected intensity patterns of high-$Q$, mid-$Q$, and low-$Q$ modes as well as the resonator geometry are shown as insets from top to bottom.
}
\label{fig1}
\end{figure}

The Bunimovich stadium is a paradigm of a fully chaotic billiard.
Therefore we expect that its resonance modes would not have high $Q$ value, since the incident angle of a ray,
initially taken even higher than the critical angle of total internal reflection,
would become easily lower than the critical angle after a few reflections, so that it escapes from the chaotic microcavity.
Figure~\ref{fig1} shows $Q$-factors of calculated modes in the single stadium shaped microcavity with $n=1.45$ when $\mathrm{Re}(kR)<30$.
As expected, the $Q$ value is not so high, the maximum $Q$ value is about 100.
Three examples of the intensity patterns are shown in Fig.~\ref{fig1}

For coupled stadium-shaped microcavities, one might naively expect the same order of $Q$ values
based on the fact that the $Q$-factor of a Fabry-Perot mode is directly proportional to the length of the cavity.
However, we find that this expectation is not always true, and there can exist abnormal high-$Q$ modes.
Figure~\ref{fig2}(a) shows $Q$-factors of the modes of coupled stadium-shaped microcavities with $n=1.45$ and $d=0.1$,
the distance between two stadium-shaped microcavities, when $\mathrm{Re}(kR)<30$.
The important feature distinct from the single stadium-shaped microcavity is
that there is a new mode group with higher $Q$, not seen in the single stadium-shaped microcavity.
Except for the high-$Q$ mode groups, the $Q$-factors of the other modes are similar to those of the single microcavity in Fig.~\ref{fig1}
in agreement with the naive expectation.
The highest $Q$-factor around $\mathrm{Re}(kR)=24.06$ is almost $1012$.
Although the total cavity size of coupled stadium-shaped microcavities is only twice as large as that of the single stadium-shaped microcavity,
the $Q$-factor is ten times higher.

\begin{figure}[tb]
\centerline{\includegraphics[width=\figsize]{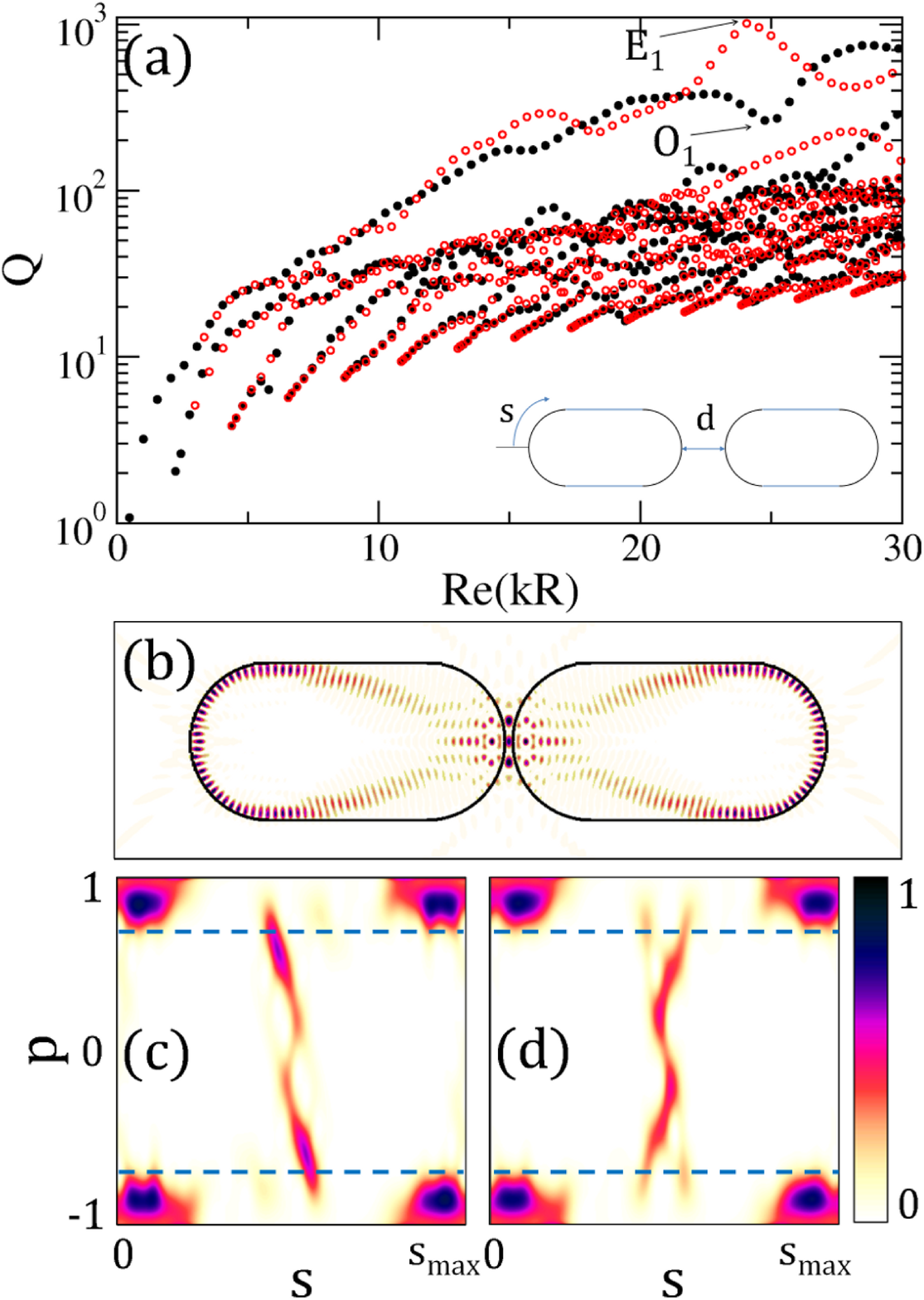}}
\caption{(color online) (a) $Q$-factors of resonant modes in coupled stadium-shaped microcavities with $n=1.45$ and $d=0.1$.
Red open and black circles represent the modes with even and odd parities about the $y$-axis, respectively.
(b) The intensity patterns of highest $Q$-modes ($E_{1}$-mode).
(c) Incident and (d) emerging generalized Husimi functions taken at the inner boundary of the left microcavity.
The blue dashed line represent the critical angle for total internal reflection.
}
\label{fig2}
\end{figure}

In order to understand the abnormal high-$Q$ modes in the coupled microcavities,
we examine the intensity pattern of the highest-$Q$ mode ($E_1$) with even parity, as  shown in Fig.~\ref{fig2}(b).
It exhibits WGM-like patterns in the circular parts of both ends and rather wide interference pattern near the interspace between two microcavities.
We call this the '{\it rounded bow tie}' mode.
The WGM-like pattern would minimize light leakage at both ends if the incident angle of the circulating wave is
greater than the critical angle $\chi_c$ of total internal reflection, $\chi_c=\arcsin (1/n)\simeq \arcsin 0.69 \simeq 43.5 ^o$.
The generalized Husimi functions \cite{Hen03} of the mode shown in Fig. 2 (c) and (d) confirm this is the case,
the incident angle at the circular part is about $\chi \simeq 58^o$ since $\sin \chi \simeq 0.85$.
This explains why the rounded bow tie mode has a high-$Q$ value.
The high-$Q$ modes of the group appear with equal spacing in $\mathrm{Re}(kR)$, like the free spectral range of WGM.
This is also the characteristics of the scarred modes showing localization along short periodic orbits of chaotic billiards \cite{Ryu13}.

\begin{figure}[tb]
\centerline{\includegraphics[width=\figsize]{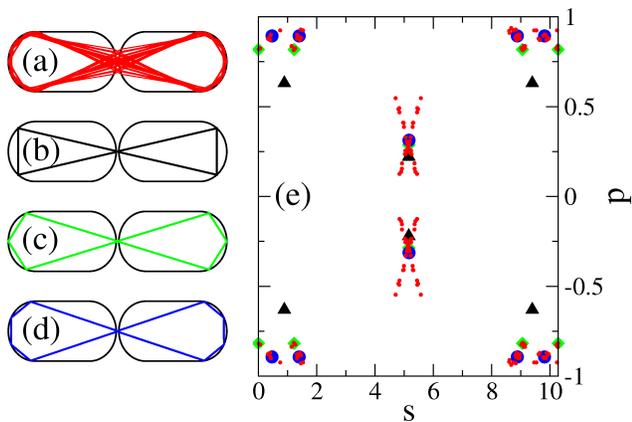}}
\caption{(color online) (a) The collection of ray trajectories corresponding to the high-$Q$ modes.
(b)-(d) The selected periodic orbits which support the high-$Q$ modes.
(e) The Poincar{\'e} surface of section for the left microcavity in the coupled stadium-shaped microcavities.
The red small circles, black triangles, green diamonds, and blue circles represent red, black, green, and blue trajectories, respectively.
}
\label{fig3}
\end{figure}

Now we consider ray dynamics to find some ray periodic orbit or ray dynamical feature explaining the rounded bow-tie mode.
Like the ray model used in the coupled microdisks \cite{Ryu10,Ryu12},
we discard the ray segment escaping from the coupled microcavities and keep the re-entering ray segment.
Then, we can get long-lived ray trajectories. These are shown in Fig. 3 (a) by the red lines.
From the close resemblance between them, it is clear that these long-lived trajectories are responsible for the high-$Q$ mode shown in Fig. 2 (b).
We also see that the two circular boundaries at the coupling region play a role of a lens focusing beams.
Thanks to this lens effect, some wave can survive for a long time so as to give rise to the WGM-like patterns at both circular ends.
In order to confirm the major role of this lens effect in forming the high-$Q$ modes, we obtain the resonance modes in a slender stadium-shaped
microcavity with $L=6R$, in the same range of $\mathrm{Re}(kR)$.
In this case, there is no dramatic enhancement of $Q$-factors.
Similar lens effects by the circular boundary have been reported in relation to the emission directionality \cite{Sha08,Lee10,Wan10,Ryu11,Shu11,Wie12}.

Using the ray model, we can find new unstable periodic orbits which contain the inside ray segments of both cavities
and the outside ray segments connecting them.
We note that there are not only unstable periodic orbits but also stable periodic orbits which forms island structures near the horizontal
bouncing ball type periodic orbit, but the stable periodic orbits are not relevant to high-Q modes.
Some unstable periodic orbits are shown in Fig. 3 (b)-(d), and, as shown in Fig. 3 (e),
these are represented in the phase space $(s,p)$, $p=\sin \chi$, by black triangles, green diamonds, and blue circles, respectively.
The long-lived ray trajectories are also denoted by the red dots in the phase space.
We can see that the long-lived trajectories are very close to the periodic orbits in Fig. 3 (c) and (d).
This closeness can be confirmed by the equal spacing of the high-$Q$ modes, $\Delta kR \simeq 0.463$, as seen in Fig. 2 (a).
If we think the scarred modes related to the periodic orbit of Fig. 3 (c),
they would show a equal spacing $\Delta kR \simeq 0.466$ from the relation $\Delta kR= 4\pi/l_{eff}$,
where $l_{eff}$ is the effective length of the periodic orbit. Similarly, the periodic orbit of Fig. 3 (d) gives a spacing $\Delta kR \simeq 0.461$.
Therefore, it is unlikely that the high-$Q$ modes are simply scarred modes of a specific periodic orbit,
but the high-$Q$ modes might have some connection with the unstable periodic orbits.

\begin{figure}[tb]
\centerline{\includegraphics[width=\figsize]{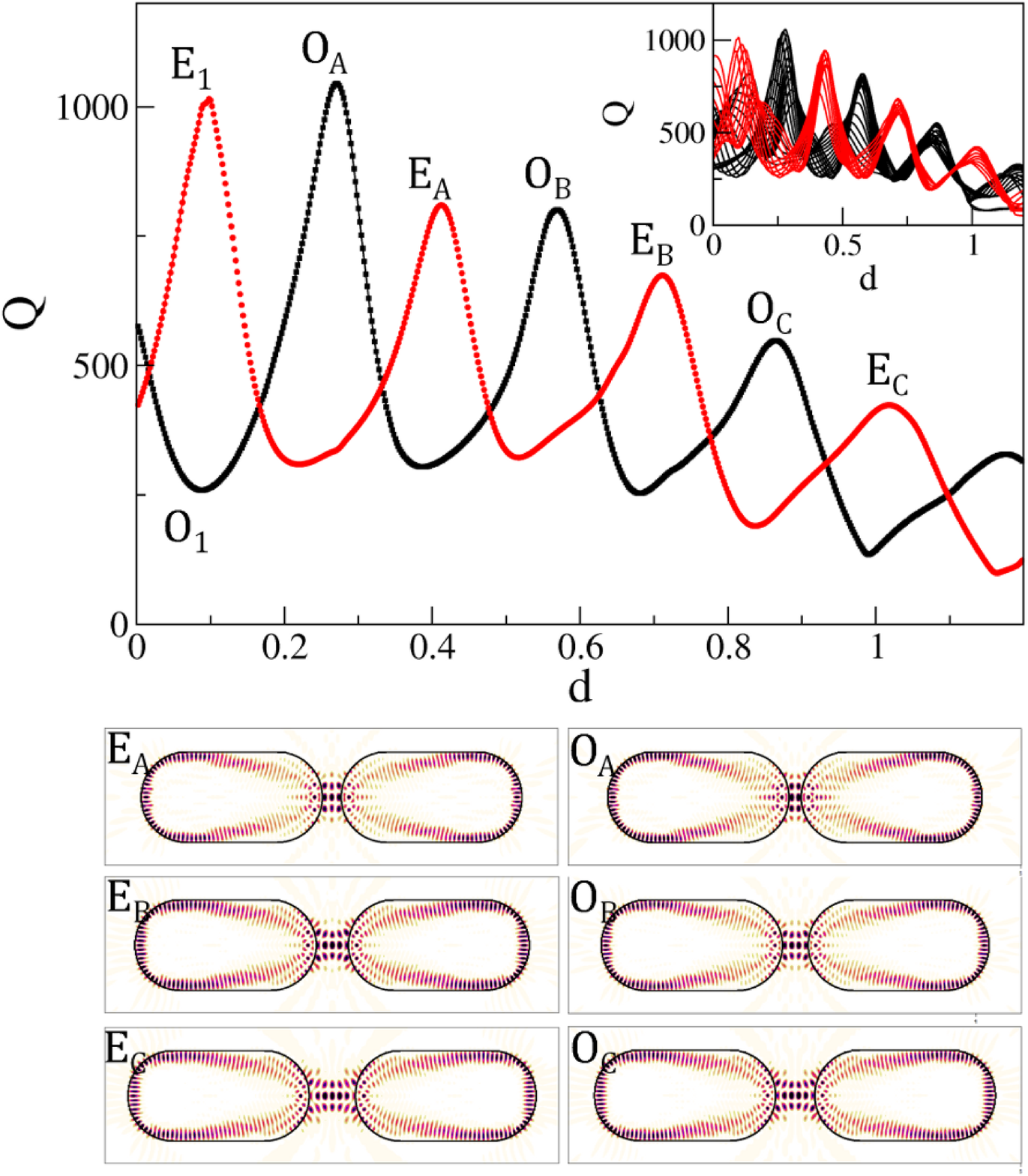}}
\caption{(color online) The $Q$-factors of $E_1$ (red squares) and $O_1$ (black circles) modes as a function of $d$.
The $E_1$ corresponds to the mode pattern of Fig.~\ref{fig2} (b).
The $E_A$($O_A$), $E_B$($O_B$), and $E_C$($O_C$) are the mode patterns at $d$
where $Q$-factors of the modes with even (odd) parity are locally highest.
(Inset) The $Q$-factors of the modes around $E_1$ and $O_1$ as a function of $d$.
}
\label{fig4}
\end{figure}

Finally, we discuss how the modes change as the distance between two stadium-shaped microcavities varies.
Figure~\ref{fig4} shows the variations of $E_1$ and $O_1$ modes which have
even and odd parity about $y$-axis, respectively.
Both modes show oscillating behaviors with almost same periods corresponding to the wavelength
$\lambda = 2 \pi / \mathrm{Re}(kR)$ but are out of phase due to the different parities.
As shown in Fig.~\ref{fig4}, the modes $E_1$, $E_A$, $E_B$, and $E_C$ have one, three, five, and seven antinodes in the interspace between two cavities,
and similarly $O_A$, $O_B$, and $O_C$ have two, four, and six antinodes according the symmetries.
The patterns near the circular boundary also change from smooth WGM-like
($E_1$ and $O_A$) into rectangle-like with angular corners ($E_C$ and $O_C$).
Modes around the $E_1$ and $O_1$ also have the high-$Q$ factors as shown in the Inset of Fig.~\ref{fig4}.
The highest $Q$-factor is always larger than about $500$ when $d$ is smaller than $0.75$.
This means that the threshold of coupled stadium-shaped microcavity lasers is much smaller than that of
a single stadium-shaped microcavity regardless of $d$ if $d$ is smaller than several wavelengths.

In summary, we have studied the characteristics of the resonant modes in coupled stadium-shaped microcavities.
We have reported the whispering gallery-like (or rounded bow tie) modes with the abnormal high-$Q$ factors appear,
which can be understood by considering the ray models.
The lens effect in the coupling area plays an important role in refocusing ray trajectories
to the vicinity of the unstable periodic orbits similar to the rounded bow tie pattern,
as a result the coupled stadium-shaped microcavities can support high $Q$ modes.
We have also discussed the variation of $Q$ factors of the modes as a function of the distance between two microcavities.
We hope this work gives some insight into coupled deformed microcavities in the viewpoint of application and theory.

This research was supported by Basic Science Research Program through the National Research Foundation of Korea (NRF) funded by the Ministry
of Education (No.2012R1A1A4A01013955 and No.2013R1A1A2065357).

\end{document}